\documentclass[pre,showpacs,citeautoscript,amsmath,twocolumn,amssymb]{revtex4}
\usepackage{setspace}

\usepackage{dcolumn}
\usepackage{bm}
\usepackage{multirow}
\usepackage{graphicx}

\begin{document}

\title{Hydrophobic interactions and hydrogen bonds in $\beta$-sheet formation.}

\author{Chitra Narayanan}
\author{Cristiano L. Dias\footnote{Author to whom correspondence should be addressed. E-mail: cld@njit.edu}}  
\affiliation{$^1$ New Jersey Institute of Technology, Physics Department, Newark, New Jersey}

\begin{abstract}
In this study, we investigate interactions of extended conformations of homodimeric peptides made of small (glycine or alanine) and large hydrophobic (valine or leucine) sidechains using all-atom molecular dynamics simulations to decipher driving forces for $\beta$-sheet formation. We make use of a periodic boundary condition setup in which individual peptides are  infinitely long and stretched. Dimers adopt $\beta$-sheet conformations at short interpeptide distances ($\xi\!\sim\!0.5$ nm) and at intermediate distances ($\sim\! 0.8$ nm), valine and leucine homodimers assume cross-$\beta$-like conformations with side chains interpenetrating each other. These two states are identified as minima in the Potential of Mean Force (PMF). While the number of interpeptide hydrogen bonds increases with decreasing interpeptide distance, the total hydrogen bond number in the system does not change significantly, suggesting that formation of $\beta$-sheet structures from extended conformations is not driven by hydrogen bonds. This is supported by an increase in electrostatic energy at short interpeptide distances. A remarkable correlation between the volume of the system and the total electrostatic energy is observed, further reinforcing the idea that excluding water in proteins comes with an enthalpic penalty. We also discuss microscopic mechanisms accounting for $\beta$-sheet formation based on computed enthalpy and entropy and we show that they are different for peptides with small and large side chains.
\end{abstract}

\maketitle
\section{Introduction}

Determining  main interactions that guide proteins to their native conformations is key to understanding protein folding, misfolding and stability \cite{DILL90}. Since the seminal work of Kauzmann \cite{KAUZ59} interactions among sidechain atoms is commonly assumed to account for tertiary contacts in proteins \cite{DILL90,LIHA97}, while backbone hydrogen bonding is associated with the stability of $\alpha$-helix and $\beta$-sheet secondary structures \cite{PAUL51,ROSE06}. Recently, hydrogen bonding has also been associated to protein misfolding and aggregation \cite{dobson2011,Maritan04,dobson08}. 
However, these interactions are difficult to quantify in aqueous solutions and are usually inferred from experiments with model compounds \cite{KAUZ59,SCHE55,BALD86}. Further, their strengths are highly dependent on the local environment within the protein. Because of this complexity it is not surprising that there is a lot of debate regarding the forces stabilizing protein conformations \cite{ROSE06,BENN06,BENN91,BENN98} and studies aiming to describe protein folding have shifted back and forth between sidechain and hydrogen bond-centered views depending on the protein property of interest \cite{Nuss2000}. This paper aims to shed light on the nature of interactions that stabilize $\beta$-sheet conformations \cite{CHEN13}.

N-Methylacetamide \cite{KLOT68, JORG89,BENT97,BUCK01} and Urea \cite{KAUZ59, SCHE55}, long used as  model systems for determining the strength of the peptide hydrogen bonds, predict negligible enthalpies for intra-peptide hydrogen bonding in water because backbone-backbone hydrogen bonds compete with water-backbone bonds to the same degree \cite{KLOT68}. However, small compounds have been argued to be bad models for the protein backbone because they lack sidechains that obstruct the formation of backbone-water hydrogen bonds. In aqueous solution, sidechains of each of the twenty amino acids have been envisaged to obstruct water differently, accounting for context-dependent hydrogen bonding \cite{BAIY94}. The strength of hydrogen bonds buried in the protein interior is observed to be as high as 7 kJ/mol per bond \cite{Martin09}, which explains the significant presence of secondary structures in globular proteins. Evidence from recent backbone mutation studies support this view \cite{DEEC04,DEEC04_JACS}. In these studies, elimination of hydrogen bonds in the dry protein core has a higher destabilizing effect (up to 5.0 kJ/mol) than elimination of hydrogen bonds at the protein's surface \cite{DEEC04, DEEC04_JACS, JICH11, JICH12, GAOJ09, PACE09}. Hydrogen bonds could therefore, play an important role in protein aggregation since about 50 \% of all bonds reside in the rigid dry core of these structures \cite{KHET00}. These different studies are part of an ongoing debate regarding the strength of backbone-backbone hydrogen bonding in water and whether they can stabilize ordered peptide structures \cite{WAYN08,ROSE06,ROSE93,HONI95}. 

Structural differences among proteins are attributed to the nature of sidechain interactions as backbone atoms are identical in all  proteins \cite{DILL90,KAUZ59,LIHA97}. In particular, the tendency of hydrophobic sidechains to be buried away from water is commonly accepted as the main driving force involved in stabilizing the native state. Hydrophobic residues are among the most conserved amino acids in protein sequences and their solubility correlates with protein stability. In particular, increased solubilities of these residues at high and low temperatures as well as high pressure have been associated with heat, cold, and pressure denaturations in proteins respectively \cite{Priv86,Priv90, RIOS00,RIOS01,DIAS08,DIAS12,DIAS10_cryobiology,HUMM98}. Strengths of hydrophobic interactions are of the order of 5 kJ/mol per --CH$_3$ group \cite{MINO94, MINO94_propensity, KIMC93, SMIT94,MERK98} and they are entropy-driven \cite{Fran45,DIAS11}. In addition to hydrophobicity, interactions between polar sidechains have also been suggested to play a major role in protein folding \cite{BENN06,BENN91,BENN98}, but this will not be addressed in this work.

In order to clarify the role of hydrophobic and hydrogen bonding interactions in protein folding and aggregation, we simulate homodimers of glycine, alanine, valine, and leucine in explicit water. To eliminate effects related to chain ends, we use periodic boundary conditions in which the carbonyl-group of the first residue is attached to the amine-group of the last residue. Hence, the environment around  each amino acid  resembles residues in the middle of strands. The energy landscape for these dimers, represented using the potential of mean force (PMF), shows well-defined global minima for all dimers at interpeptide distances corresponding to $\beta$-sheet conformations. PMFs of valine and leucine homodimers also show a second minimum at interpeptide distances of $\sim\!0.8$ nm, corresponding to conformations where sidechains of one peptide interpenetrate the space between sidechains of the neighboring peptide. Our results indicate that hydrogen bonding does not play a significant role in the formation of $\beta$-sheet structures for peptides with extended conformations. In contrast, hydrophobic interactions are shown to play a dominant role in stabilizing $\beta$-sheet conformations in peptides with large hydrophobic sidechains. We observe a striking correlation between the electrostatic energy and the total volume of the system. This supports the commonly accepted view that non-optimal packing conformations in proteins is associated with an enthalpic cost \cite{MACC07}, which can be a major rate-limiting factor in protein folding \cite{Cheung2002,RANK97}. Further, stabilizing hydrophobic interactions are shown to arise from favorable entropic contributions. These results shed light on the effect of peptide composition and its interactions with the solvent.

\section{Results}

\subsection{Potential of mean force}

\begin{figure}[h]
\centering
\includegraphics[scale=0.4]{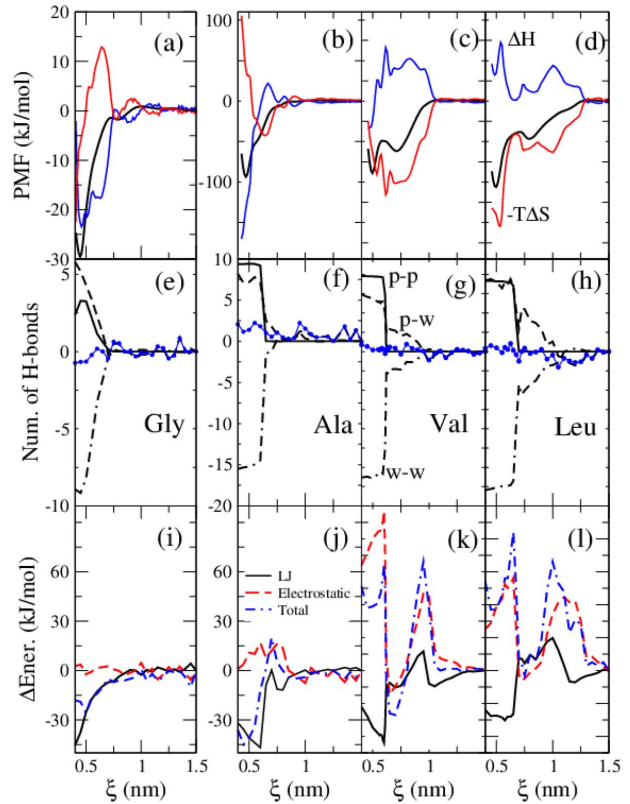}
\caption{(Panels a--d) PMFs of homopeptide dimers at 298 K (solid lines). Peptides are made of glycine (first column), alanine (second column), valine (third column), and leucine (fourth column) residues. Enthalpies and entropies are shown in dashed and dashed-dotted lines.  (Panels e--h) Number of hydrogen bonds involving peptide-peptide (solid lines), peptide-water (dashed lines), and water-water (dashed-doted lines) atoms. Net number of hydrogen bonds are shown using circles. (Panels i--l) LJ energy (solid lines), electrostatic energies (dashed lines), and total potential energy (dashed-dotted lines). Quantities computed  at $\xi=1.5$ nm are used as our reference, i.e., zero value.}
\label{PMF}
\end{figure}

Fig.~\ref{PMF} (a-d) shows the PMF at 298 K for the interaction of glycine, alanine, valine, and leucine homodimers as a function of the distance $\xi$ between the peptide centers of mass. The global minimum in the PMF is observed at $\xi = 0.5$ nm for the valine and leucine homodimers and at $\xi = 0.45$ nm in the case of the alanine homodimer. These minima can be mapped to  the $\beta$-sheet structures in which one interpeptide hydrogen bond  forms per residue (Fig.~\ref{PMF} e-h). The global minimum for the glycine homodimer occurs at $\xi = 0.40$ nm. At this distance, only one-third of the residues form hydrogen bonds. Free energies at these global minima, relative to the free energies at non-interacting distances (i.e., $\xi = 1.5$ nm)  are  -3.25 kJ/mol,  -10.0 kJ/mol, -8.25 kJ/mol, and -12.5 kJ/mol per amino acid for glycine, alanine, valine, and leucine respectively. In addition to the global minimum, PMFs of large aliphatic side-chains also show a well-defined second minimum at a distance of 0.75 nm (valine) and 0.80 nm (leucine). These second minima can be mapped to structures in which side-chains of one peptide interpenetrates the side-chains of the neighboring peptide. Previous studies using amino acid substitution experiments have shown that the free energies required to stabilize $\beta$-sheet structures cannot be attributed solely to amino acid propensities and that they also depend on the position of $\beta$-strands \cite{MINO94}. This suggests that free energies computed in this study may also be context-dependent and should therefore not be considered for their absolute values.

\subsection{Entropy and enthalpy}
To evaluate contributions of microscopic factors, we compute enthalpy and  entropy as a function of $\xi$ from the PMFs  at four temperatures (278 K, 298 K, 338 K, and 378 K) through a fit to this thermodynamic relation:
\begin{eqnarray}
\mathrm{PMF}(T,\xi) &=& \Delta H_o(\xi) - T\Delta S_o (\xi) +\\ \nonumber
& & \Delta C_{o,p} (\xi) \left[ (T - T_o) - T \log\Big(\frac{T}{T_o}\Big) \right] ,
\end{eqnarray}
where $\Delta S_o$,  $\Delta H_o$, and $\Delta C_{o,p}$ correspond, respectively, to changes in entropy, enthalpy, and heat capacity at the reference temperature  $T_o = 298$ K. Dashed-dotted and dotted lines in Fig.~\ref{PMF} (a-d) correspond to $-T_o\Delta S_o$ , and  $\Delta H_o$ respectively.

Our results show that interactions involving homodimers of glycine or alanine (Fig.~\ref{PMF}{\it a-b}) are favored by enthalpy while entropy opposes it. In contrast, interactions involving valine or leucine homodimers are disfavored by enthalpy while they are favored by entropy (Fig.~\ref{PMF}{\it c-d})---see also supplemental material \cite{sup}. These opposing behaviors indicate that adding large hydrophobic sidechains to the backbone changes the microscopic mechanism of peptide interactions. In the presence of small sidechains, peptide interactions are dominated by backbone properties that involve Lennard-Jones and electrostatic interactions between peptides atoms. Interactions involving large hydrophobic residues are dominated by the overall entropy of neighboring water molecules, which is maximized when these residues are brought close to each other. 

\subsection{Hydrogen bonds}

As the extended peptide conformations of the homodimers are brought close to each other, the number of interpeptide hydrogen bonds (solid line in Fig.~\ref{PMF} e-h) increases, reaching a maximum at distances corresponding to the global minimum of the PMF ($\xi \sim 0.5$ nm). Geometric constraints restrict the the maximum number of hydrogen bonds that can form between the backbone atoms to one per amino acid. This occurs if the two extended peptides adopt $\beta$-strand conformations \cite{PAUL51}. In the case of  alanine, valine, and leucine homodimers, this number is observed, indicating that $\beta$-sheets are formed at the global minimum of the PMF. For glycine, only one-third of all possible interpeptide hydrogen bonds are formed, suggesting that despite being stretched, these peptides do not form $\beta$-sheets.

The number of peptide-water hydrogen bonds (dashed-dotted lines) decreases when peptides of the dimer approach each other. This reflects the transfer of water molecules from the neighborhood  of the dimer towards the bulk solvent. A consequence of this transfer is an increase in the number of water-water hydrogen bonds (dashed lines) due to water added to the bulk. We note that the change in the net hydrogen bond number (blue circles) is negligible for all values of $\xi$. This results from the loss, on average, of two peptide-water hydrogen bonds and the formation of one peptide-peptide bond and one water-water hydrogen bond. This almost perfect compensation of hydrogen bonding can be attributed to two factors: (1) the ability of peptides to satisfy hydrogen bonds between polar groups in the backbone when in $\beta$-strand conformations and (2) the polar nature of the solvent which can penetrate cavities formed within the dimer  due to the small size of its molecules, and engage in hydrogen bonding with the unsatisfied polar groups of the backbone \cite{CHAN05}.

The fact that the net number of hydrogen bonds does not change for all values of $\xi$ suggests that the formation of $\beta$-sheets is not driven primarily by hydrogen bonding. This is further confirmed upon computing the electrostatic energy of the system (dashed line in Fig.~\ref{PMF} i-l). In all-atom models, hydrogen bonds stem from the sum of electrostatic terms between partial charges in amide and carbonyl groups of the backbone. If these terms played a key role in the formation of $\beta$-sheets, one would expect the electrostatic energy to be a minimum at $\xi \sim 0.5$ nm. However, this is not the case and for all systems studied, the electrostatic potential is either slightly positive (glycine and alanine) or it peaks strongly (valine and leucine) at the global minimum of the PMF.  In addition, for valine and leucine, the electrostatic contribution is the dominant term of the potential energy (dashed-dotted lines) accounting for an unfavorable potential energy to $\beta$-sheets formation. The main energetic components contributing to the potential energy for glycine and alanine peptides are the sum of all van der Waals interactions (solid lines) that are favorable for the formation of $\beta$-sheets.

\subsection{Lennard-Jones (LJ) interactions and Solvent Accessible Surface Area (SASA)}

\begin{figure}[h]
\centering
\includegraphics[scale=0.32]{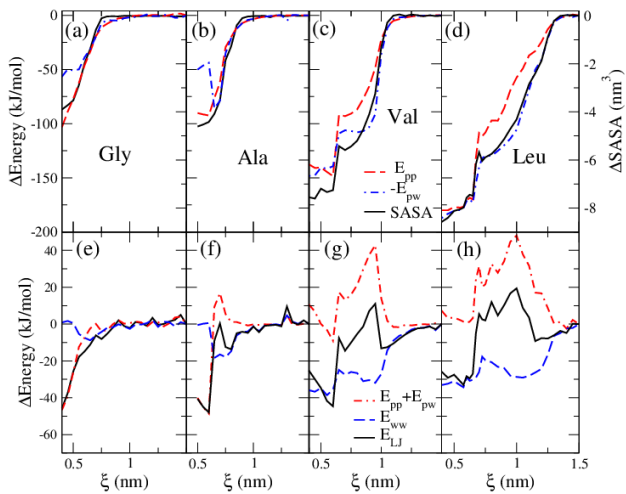}
\caption{van der Waals energy and SASA for glycine (a,e), alanine (b,f), valine (c,g), and leucine (d,h). (First row) peptide-peptide (dashed lines) and peptide-water (dashed-dotted lines) interactions correlate with SASA (solid lines). (Second row) LJ interactions involving peptide atoms (dashed-dotted lines) and water molecules solely (dashed lines) account for the total LJ energy (solid lines). Quantities computed  at $\xi=1.5$ nm are used as our reference.}
\label{LJ}
\end{figure}

Fig.~\ref{LJ} illustrates the dependence of LJ interactions and SASA on $\xi$. The top row (a-d) shows the sum of LJ interactions involving the peptide chains of the dimer E$_{\mathrm{pp}}$ (dashed lines) and the negative of LJ interactions involving simultaneously the peptide and solvent atoms -E$_{\mathrm{pw}}$ (dashed-dotted lines). Both E$_{\mathrm{pp}}$ and -E$_{\mathrm{pw}}$ decrease when peptides are brought close to each other. These interaction energies show good correlations with the SASA (solid lines).

Fig.~\ref{LJ}{\it e-h} shows the sum of the peptide-peptide (E$_{\mathrm{pp}}$) and peptide-water ( E$_{\mathrm{pw}}$) LJ interactions (dashed-dotted lines) 
These interactions show strikingly different behavior for homodimers composed of small (glycine and alanine) and larger sidechains (valine or leucine). For homodimers with small sidechains, E$_{\mathrm{pp}}$ + E$_{\mathrm{pw}}$ favors $\beta$-sheet formation while this is not observed for homodimers with large sidechains. These interactions are also unfavorable at intermediate interpeptide distances for the valine and leucine homodimers, resulting from repulsive interactions  arising due to the close proximity of sidechain atoms. Further, the LJ term involving water-water interactions, E$_{\mathrm{ww}}$, has a trend opposite to that of E$_{\mathrm{pp}}$ + E$_{\mathrm{pw}}$: it favors $\beta$-sheet formation in the valine and leucine dimers while it plays no role for the glycine and alanine homodimers.

In the case of valine and leucine dimers, these results are consistent with a picture for $\beta$-sheet formation in which water molecules found in the space between sidechains are released into bulk water accounting for a decrease in E$_{\mathrm{ww}}$. In this process, the number of peptide-peptide contacts increases corresponding to a  decrease in E$_{\mathrm{pp}}$ while the number of water-peptide contacts decreases, corresponding to an increase in E$_{\mathrm{pw}}$. These opposing behaviors in E$_{\mathrm{pp}}$ and E$_{\mathrm{pw}}$ account for the observed negligible change in E$_{\mathrm{pp}}$ + E$_{\mathrm{pw}}$ observed at $\xi \sim$ 0.5 nm (i.e., at $\beta$-sheet structures). For homodimers with small sidechains, water molecules can partially permeate the space between sidechains when peptides are in $\beta$-sheet conformations \cite{DIAS11}, resulting in an increase in peptide-peptide contacts while retaining some peptide-water contacts. As a result, E$_{\mathrm{pp}} $ + E$_{\mathrm{pw}}$ is  favorable while  E$_{\mathrm{ww}}$ does  not change significantly.

\subsection{Electrostatic energy and cavities}

\begin{figure}[h] 
\centering
\includegraphics[scale=0.32]{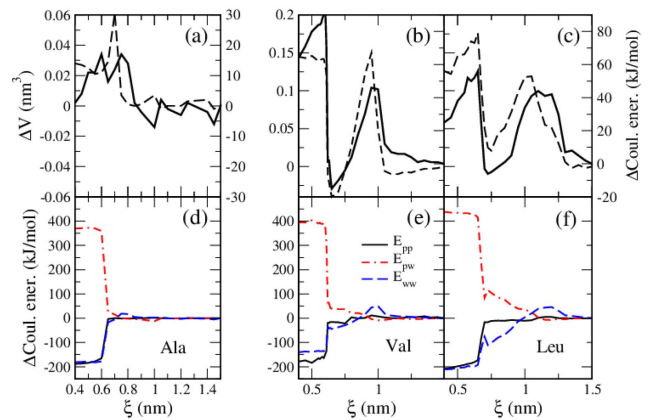}
\caption{(a-c) Electrostatic energy and total volume for alanine, valine, and leucine respectively. (d-f) Electrostatic energy contributions from peptide-peptide (solid lines), peptide-water (dashed-dotted lines) and water-water interactions (dotted lines) for alanine, valine, and leucine respectively. Quantities computed  at $\xi=1.5$ nm are used as our reference.}
\label{Volume}
\end{figure}

Fig.~\ref{Volume} (a-c) shows the dependence of the electrostatic energy (solid lines) and total volume (dashed lines) on $\xi$. These quantities are not shown for the glycine homodimer because they do not vary significantly as a function of $\xi$. $\beta$-sheet formation is associated with an increase in the volume in the order of $\Delta V = 0.02$ nm$^3$ for the alanine homodimer (panel a) and  $\sim\!\!0.15 \mathrm{nm}^3$ for the valine and leucine homodimers (panels b and c). The formation of interpenetrating sidechain configurations also leads to a similar increase in volume ($\sim\!\!0.15 \mathrm{nm}^3$) for valine and leucine homodimers. A minimum in the volume is observed at distances of $\xi\!\!\sim$0.75 nm at which interpenetrating sidechain configurations are highly compact. Furthermore, an excellent correlation is observed between the electrostatic energy and the total volume (Fig.~\ref{Volume} (a-c). To decipher the origin of this correlation, we show peptide-peptide, peptide-water, and water-water contributions to the electrostatic energy in Fig.~\ref{Volume} (d-f).

At interpeptide distances corresponding to $\beta$-sheets, magnitudes of favorable peptide-peptide (solid lines) and water-water (dashed lines) electrostatic energies are almost identical. When added together their sum is slightly less than the magnitude of unfavorable peptide-water (dashed-dotted lines) electrostatic energy. This   accounts for the net unfavorable electrostatic energy associated with $\beta$-sheet formation (solid lines in panels {\it a-c}). If we stipulate that hydrogen bond strengths are not strongly dependent on the atomic species interacting in water, then hydrogen bonds only account for a negligible change in the electrostatic energy. In this case, the unfavorable electrostatic energy for $\beta$-sheet formation can only stem from the loss of sidechain-water electrostatic interactions, as water molecules in between sidechains of neighboring peptides are transferred into the bulk. Accordingly, we observe an increase in peptide-water electrostatic energies (dashed-dot lines) with sidechain size at $\xi = 0.5$ nm (385 kJ/mol for alanine, 400 kJ/mol for valine, and 450 kJ/mol for leucine). Furthermore, this loss of sidechain-water interactions accounts for an increase in the volume of the system, as cavities are left in the space between sidechains. We note that sidechain-sidechain electrostatic interactions between non-polar residues are negligible.

\begin{figure}[h]
\includegraphics[scale=0.4, angle=0]{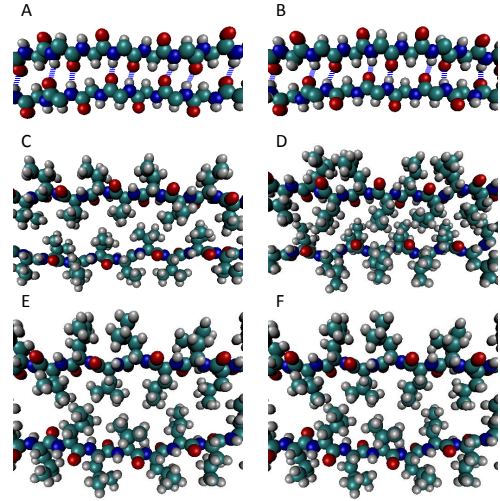}
\caption{Representative conformations of valine (column 1) and leucine (column 2) homodimers at interpeptide distances of 0.5 (a,b), 0.75 (c,d), and 1.0  nm (e,f) respectively. Sidechains in a-b are hidden for clarity, while blue lines represent the inter-strand hydrogen bonds.}
\label{DOW}
\end{figure}

The observation of large volume and large electrostatic energy at $\xi = 1$ nm (panels {\it b} and {\it c}) correlates with unfavorable water-water electrostatic energy (dashed lines in panels {\it e-f}). They correspond to loosely packed interpenetrating sidechain configurations with large cavities between backbones. These structures have a large SASA (Fig.~\ref{LJ}), in addition to a large number of shell-water with less favorable water-water interactions compared to bulk water. Characteristic conformations at interpeptide distances of 0.5, 0.75 and 1.0 nm are shown in Fig.~\ref{DOW}.

As $\xi$ decreases, interpenetrating configurations become more compact as sidechains are packed against each other filling the cavities between backbones (Fig.~\ref{DOW}{\it c,d}). This accounts for a decrease in both SASA and the number of shell water. At $\xi =0.75$ nm, the SASA is a minimum for interpenetrating sidechain configurations (see Fig.~\ref{LJ}),corresponding to a minimum in the water-water electrostatic energy (Fig.~\ref{Volume}{\it b,c}). Note that for interpenetrating configurations, peptide-water electrostatic energy increases with decreasing $\xi$ (dashed-doted lines). This can be rationalized in terms of the reduction in the number of shell water.

\hspace{-5cm}

\section{Discussion and conclusions}

In this study, we have simulated homodimers made of glycine, alanine, valine and  leucine to determine the role of hydrophobic and hydrogen bonding interactions in the formation and stability of $\beta$-sheet conformations. Our results indicate that hydrogen bonding does not play a significant role in the formation of $\beta$-sheets for peptides in extended conformations: the net number of hydrogen bonds in the system does not change as a function of peptide dimer distance and electrostatic energies peak at distances corresponding to $\beta$ structures. If hydrogen bonds were  dominant interactions stabilizing $\beta$-sheets, electrostatic energies would be a minimum. One limitation of this work is that we consider non-interacting configurations (large values of $\xi$) as extended conformations, instead of the unfolded conformations of proteins. While these extended conformations are part of the random coil ensemble, our structures do not include individual configurations that might have unsatisfied hydrogen bonds, which could contribute hydrogen bonds during protein folding.

The results of this work can be applied to explain the phenomenon of cold denaturation. The most accepted explanation for this phenomenon relates cold denaturation to decreasing stabilities of hydrophobic interactions  upon cooling. This leads to greater exposure of the dry protein core to water, accounting for unfolding at low temperatures. Recent studies have shown cold denaturation in $\beta$-hairpin peptides which have a limited buried hydrophobic core \cite{DYER05}. Simulations with implicit solvent have attributed this phenomenon to reduced backbone hydrogen bond stability upon cooling \cite{SHAO13}. In our simulations, the stability of $\beta$-sheets made of valine or leucine residues decreases with decreasing temperature, accounting for the entropy-driven behavior shown in Fig.~\ref{PMF}. This is a requirement for cold denaturation to occur.
Our simulations therefore support the conventional explanation for cold denaturation of $\beta$-sheet peptides containing large hydrophobic sidechains.  In contrast, the stability of $\beta$-sheet peptides made of alanine amino acids  in our simulations, increased with decreasing temperature, accounting for the enthalpy-driven behavior shown in Fig.~\ref{PMF}. This suggests that cold denaturation may not occur for alanine peptides.

$\beta$-sheets are core structures in fibrillar aggregates of amyloid peptides \cite{NELS05}, which can vary significantly in amino acid composition. Based on our simulations, we speculate that fibril formation is enthalpy-driven for sequences composed of short sidechains, like the  residues 113-120 (AGAAAAGA) of Syrian hamster prion protein \cite{DAID04}. Contributions to this enthalpy arise primarily from van der Waals interactions between peptide atoms. On the other hand, fibril formations in sequences made predominantly of large hydrophobic sidechains, for example residues 68-78 (GAVVTGVTAVA) in human $\alpha$-synuclein \cite{DAVI98,TREX09}, are entropy-driven, as observed for our leucine and valine homodimers. Microscopically, water molecules are responsible for this entropic interaction \cite{Fran45,DIAS11_hydrophobic}. Hence, we expect that changes in the solvent environment will have a stronger effect on fibrillar structures of peptides containing large hydrophobic residues compared to alanine-based peptides. 

In summary, we have used homodimeric peptide systems to clarify the role of hydrophobic and hydrogen bonding interactions in the stability of $\beta$-sheet conformations in peptides. Free energy of the  interactions between the homodimeric peptides  is calculated using umbrella sampling with the interpeptide distance as an order parameter. We show that the free energies of the four systems studied are characterized by two minima corresponding to $\beta$-sheet structures and peptide conformations with interpenetrating sidechain configurations. We determined the energetic contributions to these minima and found that (1) interpeptide hydrogen bonds do not contribute significantly to the stability of sheets; (2) electrostatic energies correlate with the volume of the system; (3)  $\beta$-sheet formation in peptides with large hydrophobic sidechains (valine and leucine) is entropy-driven while they are enthalpy-driven in peptides with small sidechains (alanine and glycine). These results shed light and contribute to answering longstanding questions related to roles played by hydrogen bonds and hydrophobic interactions in the stability of protein structures. We are confident that the conceptual framework and methodology developed here will facilitate elucidation of these fundamental questions through further efforts in theory and experiment.

\section{Materials and Methods}

\begin{figure}
\includegraphics[scale=0.8, angle=0]{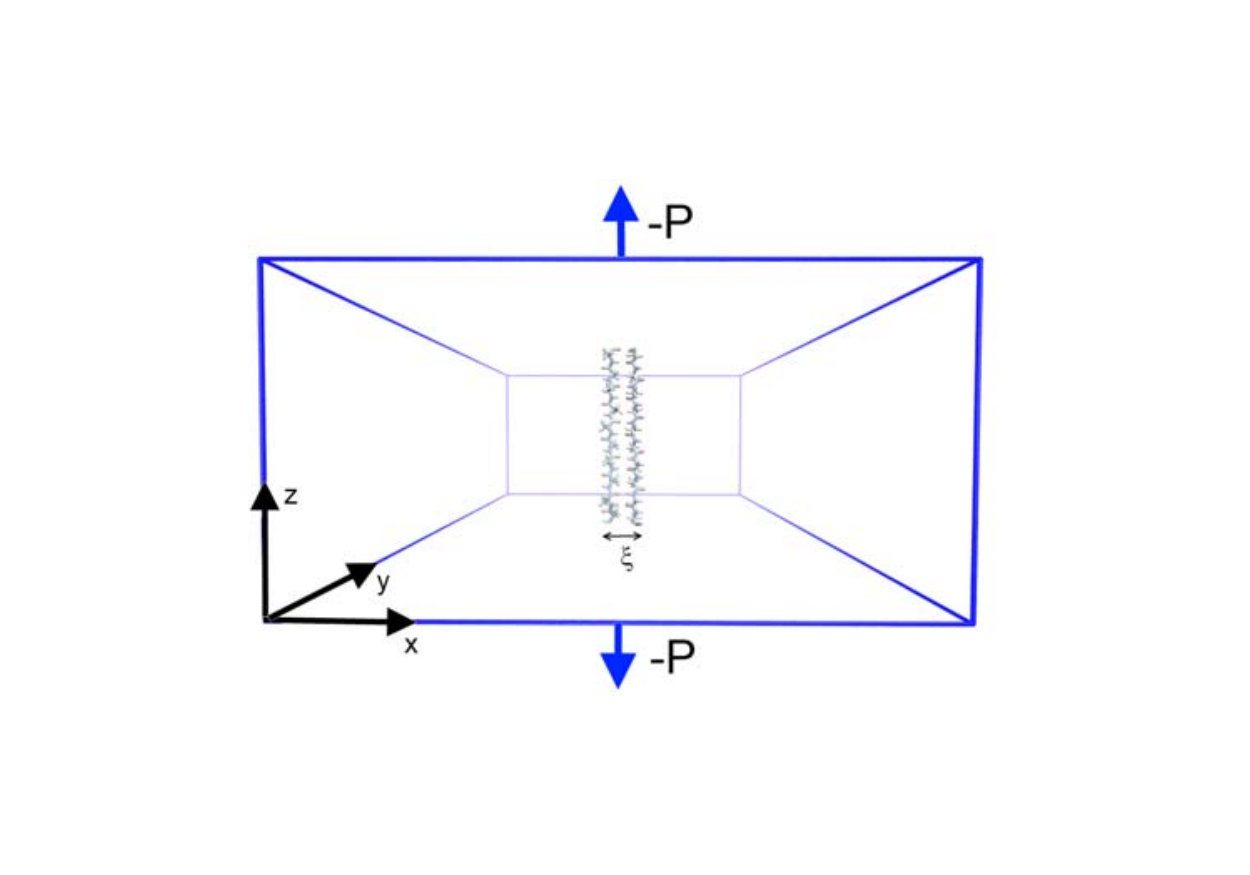}
\vspace{-2cm}
\caption{Schematic representation of the simulation box with the two peptides. $\xi$ is the order parameter corresponding to the distance between the centers of mass of the two peptides. Peptides are made of ten amino acids with residues ``1'' and ``10'' attached ``covalently'' to each other through the boundary in the z-direction, forming an infinite chain.}
\label{setup}
\end{figure}

In this work, we simulate two infinite homopeptides at different distances $\xi$ between their centers of mass (Fig.~\ref{setup}). The two peptides are placed in anti-parallel orientations resulting in $\beta$-sheet formation at small values of $\xi$. Four homodimeric peptide systems composed of glycine, alanine, valine and leucine residues respectively were used. Each of the homodimeric Peptides, which are ten amino acids long, are made infinite through periodic boundary in the z-direction: the carbonyl-group of residue 1 is attached to the amine-group of residue 10. The use of infinite chains eliminates effects from chain ends and all amino acids become equivalent, resembling amino acids in the middle of a strand. A potential constraint of this setup is that it does not allow the formation of twists which have been shown to affect the stability of $\beta$-sheets through increased sidechain interactions \cite{CHOU83,CHOU83b}. However, this limitation does not affect our conclusion: if twists were added to our model, sidechain interactions would play a more important in the stability of $\beta$-sheets with respect to hydrogen bonds than the one computed here.

Peptides are immersed in a box of $\sim$5,500 water molecules (TIP3P) and a  pressure is applied along {\it z}-direction (main axis of the peptides) to keep the box from collapsing. The magnitude of this pressure is chosen to ensure an average peptide length of is 3.5 nm. A pressure of 1 atm is applied along {\it x} and {\it y} directions to account for water density at ambient pressure. Simulations are carried out using GROMACS and CHARMM27 forcefield. Temperature and Pressure were controlled using the velocity-rescale thermostat ($\tau_{\mathrm{T}}$ = 1 ps) and the Parrinello--Rahman barostat ($\tau_{\mathrm{P}}$ = 1 ps), respectively. Simulations were performed with a time step of 2 fs and the neighbor list was updated every 10 steps. Electrostatics were treated by the Smooth Particle Mesh Ewald with a grid spacing of 0.13 nm and a 1.3 nm real-space cutoff \cite{BJEL10}. We use umbrella sampling with a spring constant of 5,000 kJ/mol to compute the properties of the system at different windows of $\xi$ ranging from 0.4~nm~ to 1.7~nm. Windows are separated from each other by 0.05~nm. Simulations lasted at least 100 ns in each window. PMFs of homodimers were computed using the Weighted Histogram Analysis Method (WHAM). 

To define hydrogen bonds we employ a commonly used geometrical definition in which these bonds are formed when the distance between donor (D) and acceptor (A) is smaller than 0.4 nm and the angle H-D-A  is smaller than 30$^o$.  We used g\_hbond provided in the software package GROMACS for this calculation. In all figures, computed values of hydrogen bonds and energies at $\xi=1.5$ nm were used as our zero reference. In the calculation of quantities involving solute--solvent and solvent--solvent atoms, all solvent (water) molecules were taken into account. However, since we only report differences with respect to quantities computed at $\xi = 1.5$ nm, the effect of bulk water is averaged out. Using all water molecules resolves the problem of having to define a cut-off for solvent molecules  in the calculation of solute--solvent and solvent--solvent properties.

\section{Acknowledgment}
I would like to thank Markus Miettinen for providing us with the topology files of periodic peptides, Roland Netz for insightful discussions and for hosting my stay at Freie Universit\"at Berlin when I started running simulations, and Normand Mousseau for hosting my stay at the Universit\'e de Montr\'eal when the ideas of this project were taking shape. This project was partially funded by the Volkswagen Foundation. I would like to thank Compute Canada for computational resources.




\end{document}